\documentclass[8.5pt,twoside,twocolumn]{article}
\usepackage[super,sort&compress,comma]{natbib}
\usepackage{mhchem}
\usepackage{times,mathptmx}
\usepackage{sectsty}
\usepackage{balance}
\usepackage{amssymb}

\usepackage{graphicx} 
\usepackage{lastpage}
\usepackage[format=plain,justification=raggedright,singlelinecheck=false,font=small,labelfont=bf,labelsep=space]{caption}
\usepackage{fancyhdr}
\begin{document}

\thispagestyle{plain}

\makeatletter
\def\subsubsection{\@startsection{subsubsection}{3}{10pt}{-1.25ex plus -1ex minus -.1ex}{0ex plus 0ex}{\normalsize\bf}}
\def\paragraph{\@startsection{paragraph}{4}{10pt}{-1.25ex plus -1ex minus -.1ex}{0ex plus 0ex}{\normalsize\textit}}
\renewcommand\@biblabel[1]{#1}
\renewcommand\@makefntext[1]%
{\noindent\makebox[0pt][r]{\@thefnmark\,}#1}
\makeatother
\renewcommand{\figurename}{\small{Fig.}~}
\sectionfont{\large}
\subsectionfont{\normalsize}

\renewcommand{\headrulewidth}{1pt}
\renewcommand{\footrulewidth}{1pt}
\setlength{\arrayrulewidth}{1pt}
\setlength{\columnsep}{6.5mm}
\setlength\bibsep{1pt}

\twocolumn[
  \begin{@twocolumnfalse}
\noindent\LARGE{\textbf{The phase diagram of water from quantum simulations}}
\vspace{0.6cm}

\noindent\large{\textbf{Carl McBride,\textit{$^{a}$} Eva G. Noya,\textit{$^{b}$} Juan L. Aragones\textit{$^{a}$}, Maria M. Conde\textit{$^{a}$}, and Carlos Vega\textit{$^{a\ddag}$}}}\vspace{0.5cm}

\noindent\textit{\small{\textbf{Received 26th March  2012, Accepted 22nd May 2012\newline
First published on the web 22nd May 2012}}}

\noindent \textbf{\small{DOI: 10.1039/C2CP40962C}}
\vspace{0.6cm}

\noindent \normalsize{

The phase diagram of water has been calculated for the TIP4PQ/2005 model, an empirical rigid non-polarisable model.
The path integral Monte Carlo technique was used, permitting the incorporation of nuclear quantum effects.
The coexistence lines were traced out using the Gibbs-Duhem integration method, once having calculated the
free energies of the liquid and solid phases in the quantum limit, which were obtained via thermodynamic integration from the classical value by
scaling the mass of the water molecule.
The resulting phase diagram is qualitatively correct, being displaced to lower temperatures by 15-20K.
It is found that the influence of nuclear quantum effects are correlated to the tetrahedral order parameter.

}

\vspace{0.5cm}
 \end{@twocolumnfalse}
  ]

\footnotetext{\ddag~carlos@ender.quim.ucm.es}
\footnotetext{\textit{$^{a}$~Departamento de Qu\'{\i}mica F\'{\i}sica, Facultad de Ciencias Qu\'{\i}micas, Universidad Complutense de Madrid, 28040 Madrid, Spain}}
\footnotetext{\textit{$^{b}$~Instituto de Qu\'{\i}mica F\'{\i}sica Rocasolano, Consejo Superior de Investigaciones Cient\'{\i}ficas, CSIC, Calle Serrano 119, 28006 Madrid, Spain}}

\section{INTRODUCTION}
Ever since the monumental work undertaken by Bridgman in 1912 \cite{PAAAS_1912_XLVII_13_441}
there has been intense and continued interest in the phase diagram of water \cite{PRL_2009_103_105701,PCCP_2011_13_18468}. 
The prediction of the phase diagram serves as a severe test
for any model of water \cite{JCP_1982_76_00650,JCP_1984_81_04087}. 
Although the first computer simulations of water were performed in 1969 by Barker and Watts \cite{CPL_1969_03_0144} and 1971 by Rahman and Stillinger \cite{JCP_1971_55_03336},
the calculation of the complete phase diagram was only recently undertaken, using the  classical models TIP4P and SPC/E \cite{PRL_2004_92_255701}.
Although the TIP4P model provided a  qualitatively correct phase diagram, there was room for improvement (i.e. the melting
point of ice I$_{\mathrm{h}}$ was situated at around 230K). 
 Consequently  a new re-parameterisation, named TIP4P/2005, was proposed \cite{JCP_2005_123_234505} 
leading to a satisfactory description of a number of properties of water \cite{PCCP_2011_13_19663,PRL_2011_107_155702}. 
The TIP4P/2005 model is a rigid non-polarisable model designed for classical simulations. 
In an indirect fashion  TIP4P/2005 implicitly incorporates nuclear quantum effects, at 
least at moderate to high temperatures. However, the model fails when it comes to  describing
the equation of state at low temperatures \cite{JCP_2009_131_024506} or  $C_p$ \cite{JCP_2010_132_046101}. The origin of 
the failure is the use of classical simulations to describe the properties of water. 
Quantum effects are present in water \cite{JACS_2005_127_05246,JCP_2008_128_074506,PRL_2008_101_017801,JPCB_2009_113_05702,JCP_2010_133_124104,JCP_2011_135_224111} even at ``high'' temperatures, 
due to the particularly small moment of rotational inertia, engendered by the low mass of hydrogen, in conjunction with 
the relatively high strength of the intermolecular hydrogen bonds. 

Nuclear quantum effects can be incorporated into condensed matter simulations via the path integral
technique proposed by Feynman \cite{RMP_1948_20_00367} (for an excellent review see \cite{NATO_ASI_C_293_0155_photocopy}).  
Barker \cite{JCP_1979_70_02914} and
Chandler and Wolynes \cite{JCP_1981_74_04078} showed that the formalism
of Feynman is equivalent, or ``isomorphic'', to performing classical simulations of a modified system
where each molecule is replaced by a polymeric ring composed of $P$ beads.
The TIP4P/2005 model, successful for classical simulations \cite{JACS_2011_133_01391,JACS_2011_133_06809}, was recently adjusted for use in such quantum simulations 
(the charge located on the hydrogen  atom was increased by $0.02e$ so as to maintain the same internal energy in a quantum simulation as the TIP4P/2005 model in a classical simulation)
becoming the TIP4PQ/2005 model \cite{JCP_2009_131_024506}. This new variant of TIP4P/2005 has been successful in describing the
temperature of maximum density \cite{JCP_2009_131_124518} of water and heat capacities \cite{JCP_2010_132_046101}.
It is for this model that we calculate the phase diagram.

\section{METHODS}
\subsection{Path integral Monte Carlo}
The partition function, $Q_{NpT}$,  for a system of $N$ rigid molecules in the $NpT$ ensemble is given (except for an arbitrary pre-factor 
that renders $Q_{NpT}$ dimensionless) by
$Q_{NpT}= \int  \exp(-\beta pV) Q_{NVT} ~{\mathrm{d}}V$.
In the $NVT$ ensemble $Q_{NVT}$ is given by:
\begin{eqnarray}
\label{partition_function}
&&\frac{1}{N!}\left( \frac{MP}{2\pi\beta \hbar^{2} } \right)^{\frac{3NP}{2}} 
\int \prod_{i=1}^N \prod_{t=1}^{P} \rho_{\mathrm{rot},i}^{t,t+1} \left( \frac{\beta}{P}, {\bf \omega}_i^t {\bf \omega}_i^{t+1}  \right)  ~{\mathrm{d}}\mathbf{r}_i^t ~{\mathrm{d}}{\bf \omega}_i^t    \nonumber \\
&& \times 
\exp \left( - \frac{MP}{2\beta \hbar^2} \sum_{i=1}^N \sum_{t=1}^P\left( \mathbf{r}_i^t - \mathbf{r}_i^{t+1}\right)^2 -\frac{\beta}{P} \sum_{t=1}^P U^{t}\right) 
\end{eqnarray}
where $P$ is the number of Trotter slices or ``replicas"   through which  nuclear quantum effects are introduced (for $P=1$ the simulations become classical). 
Each replica, $t$, of molecule
$i$ interacts with the replicas with the same index
$t$ of the remaining particles via the
inter-molecular potential $U$,
and interacts with replicas $t-1$ and $t+1$ of the same molecule $i$ through a harmonic
potential (connecting the centre of mass of the replicas) 
whose coupling parameter depends on the mass of the molecules ($M$)
and on the temperature ($\beta=1/k_BT$),
and through a term ($\rho_{\mathrm{rot},i}^{t,t+1}$), named the rotational propagator, that incorporates the quantisation
of the rotation and which depends on the relative orientation
of replicas $t$ and $t+1$. Pioneering work was undertaken by Wallqvist and Berne \cite{CPL_1985_117_0214}, and by  Rossky and co-workers \cite{CPL_1984_103_0357}
who used an approximate expression for the rotational propagator of an asymmetric top (i.e. water). 
Another technique is that of the stereographic projection path integral \cite{JCP_2008_128_204107} which has been used to study TIP4P clusters \cite{JCP_2009_131_184508}.
In 1996 M\"user and Berne \cite{PRL_1996_77_002638} provided an expression of the rotational propagator 
for spherical and symmetric top  molecules.  Quite recently the authors have extended the expression of M\"user and Berne
to the case of an asymmetric top \cite{JCP_2011_134_054117}, which is the case of water. 
The propagator is a function of the relative Euler angles between two contiguous beads and of $PT$. 
The internal energy \emph{E} can be calculated through the derivative of the logarithm of $Q_{NVT}$ with respect 
to $\beta$, and is given by the sum of the kinetic and potential energy terms, $E = K_{\mathrm{translational}} + K_{\mathrm{rotational}} + U = K  + U $.
A more complete account concerning path integral Monte Carlo simulations of rigid rotors, and their application to water,
can be found in the article by Noya et al. \cite{MP_2011_109_0149}.
\subsection{Phase diagram calculation}
The determination of the phase diagram of the quantum system is undertaken in several steps.
First the classical phase diagram of the TIP4PQ/2005 model was calculated.
To do this, for each solid phase a reference thermodynamic state is chosen and the free energy of the 
classical system is determined 
using either the Einstein crystal \cite{JCP_1984_81_03188} or the Einstein molecule methodologies \cite{JCP_2008_129_104704}. 
For the fluid phase the free energy of the classical system is determined at a reference state
by transforming the TIP4PQ/2005 model into the Lennard-Jones model, for which the free energy is well known \cite{MP_1993_78_0591}.  
The free energy of the classical system under distinct thermodynamic conditions can be  obtained
via thermodynamic integration. This permits one to determine an initial  coexistence point of the classical system 
for each phase transition
by imposing the usual condition of equal chemical potential for a given $T$ and $p$. 
Gibbs-Duhem simulations \cite{JCP_1993_98_04149} are then performed to trace out the complete phase diagram.
The procedure has been described in detail in \cite{JPCM_2008_20_153101}. At the end of this first step
the phase diagram of the classical system is known. 

In the second step the chemical potential of the quantum system is determined at a reference thermodynamic
state, again for each phase of interest. It is worth  describing this procedure in some detail. 
Let us define the excess quantum free energy as the free energy difference
between the quantum system and its classical counterpart at the same $T$ and $p$:
\begin{equation}
 G^{ex,Q} =  G - G^{\mathrm{classical}}  
\end{equation}
Thus the free energy of the quantum system,  $G$, can be obtained if the classical and excess contributions
are known.
The free energy of the classical system was determined
in the first step, so we shall now focus on the evaluation of $G^{ex,Q}$.
One defines a parameter, $\lambda$, whose purpose is to scale  the
mass of the atoms of the molecule of water such that:
$m_{O}=\lambda m_{O,0} = m_{O,0}/\lambda'$ and 
$m_{H}=\lambda m_{H,0} = m_{H,0}/\lambda'$, 
where $m_{O,0}$ and $m_{H,0}$ are the masses of O and H in the molecule of water
and where   $\lambda'=1/\lambda$. 
Thus one can can slide  from the quantum limit 
(for which $\lambda'=$1) to the
classical limit (for which $\lambda'=$0) by simply changing the $\lambda'$ 
parameter.
From the relationship
$G= -kT \ln Q_{NpT}$ the derivative of the free energy with respect to  $\lambda'$ 
can be calculated \cite{libro_free_energy}, obtaining:
\begin{equation}
\label{derivada}
\frac{\partial (G/NkT)}{\partial \lambda'} = \frac{1}{\lambda'} 
\left\langle \frac{K}{NkT} \right\rangle \\
\end{equation}
This is due to the fact that when the mass of all atoms of the molecule
are scaled by a factor $1/\lambda'$ the total mass, $M$, is also scaled by a factor $1/\lambda'$. 
The same is true for the eigenvalues of the inertia tensor and thus  the energies of
the asymmetric top appearing in the rotational propagator are also scaled by a factor $\lambda'$.
The average of the value in the angled brackets should be performed for the value of $\lambda'$ of interest. 
By using Eq. \ref{derivada} in conjunction with  the fact that the total kinetic energy (with the translational
and rotational contributions) is $3NkT$ for the classical and for the quantum system in the limit
of infinitely heavy molecules, it can be shown that the chemical potential of the quantum system $\mu$ 
can obtained from the expression:
\begin{equation}
\label{calculating_G_exc}
\frac{\mu}{kT}=  \frac{G^{\mathrm{classical}}}{NkT} +
\int_0^1 \frac{1}{\lambda' } \left[ \left\langle \frac{K}{NkT} \right\rangle-
3  \right] d\lambda'    \\
\end{equation}
To determine the integral of Eq. \ref{calculating_G_exc} (i.e.  $G^{ex,Q}/NkT$) 
it is sufficient to perform simulations at  decreasing values of
$\lambda'$, to determine the integrand for each considered value of $\lambda'$ 
and subsequently implement a numerical procedure to estimate the value of the integral.   
It follows from Eq. \ref{calculating_G_exc} that the difference in chemical potential in the quantum
system between two phases,  $\Delta \mu =  \mu_B - \mu_A$,  can be obtained as:
\begin{eqnarray}
\frac{\Delta \mu} {kT} =   \frac{\Delta \mu^{\mathrm{classical}} }{kT} 
+ \int_{0}^{1}\frac{1}{\lambda'} 
\left[ \left\langle \frac{K_{B}}{NkT} \right\rangle -  \left\langle \frac{K_{A}}{NkT} \right\rangle 
 \right] \mathrm{d}\lambda' 
\label{eq_guay}
\end{eqnarray}
This expression states that the difference in chemical potential between two phases is simply the value of the difference in the classical system
plus a correction term that accounts for the difference in the quantum excess free energies. 
Thus, after the second step one knows either the
chemical potential of the quantum system at a reference state, or similarly the difference in chemical
potential between two phases, again  at a reference state. 

The third step in the determination of the phase diagram of the quantum system requires 
the determination of one initial  coexistence point  for each coexistence line.
By using thermodynamic integration \cite{JPCM_2008_20_153101},
the free energy of each phase of the quantum system is determined as a function of $T$ and $p$.
This provides the location of at least one coexistence point between each pair of phases by imposing the condition
of identical chemical potential, $p$ and $T$ between the two phases. 

The fourth and final step is the tracing out  of the complete coexistence lines thus yielding the
phase diagram. This is done by using the Gibbs-Duhem simulations, starting from the initial
coexistence point determined at the end of the third step. 
\subsection{Simulation details}
The expression of the rotational propagator is given in \cite{JCP_2011_134_054117}. The propagator
is composed of an infinite sum over the energy levels of the
free asymmetric top rotor.
In practice the summations are truncated;
we adopted the criterion that the
propagator had converged when the absolute difference between the value of the propagator  for
two consecutive values of $J$ (normalised so that $\rho (0,0,0)=1$ for both values of $J$)
is less than $10^{-6}$ per point.
A grid of one degree for each Euler angle was used, and results for intermediate  angles were obtained by interpolation.
Simulations consisted of 360 molecules for liquid water, 432 molecules for ices I$_{\mathrm {h}}$ and II,
324 molecules for ice III, 504 for ice V and 360 for ice VI.
The algorithm of Buch  {\it et al.} \cite{JPCB_1998_102_08641} was used to obtain a proton disordered
configuration of ices  I$_{\mathrm {h}}$, III, V and VI simultaneously having zero dipole moment and at the same time satisfying the Bernal-Fowler
rules \cite{JCP_1933_01_00515,JACS_1935_57_02680}. For ice III  we additionally imposed the condition that the selected
proton disordered configuration presented an internal energy
that lies in the centre of
the energy distribution shown in Figure 2 of \cite{JPC_A_2011_115_005745}.
Direct I$_{\mathrm{h}}$-fluid coexistence simulations used about 1000 molecules for the classical system and
about 600 for the quantum one.
Free energies of the classical system were obtained using the Einstein molecule
  methodology for a given proton disordered configuration and subsequently adding the
 Pauling \cite{JACS_1935_57_02680}
  entropy contribution ($-RT \ln(3/2)$). The methodology has been described
  in detail elsewhere \cite{JPCM_2008_20_153101}.
  The Lennard-Jones part of the potential was truncated at 8.5{\AA}
  and long ranged corrections were added. Coulombic interactions were treated using
  Ewald sums.  For the solid phase anisotropic $NpT$ Monte Carlo simulations were performed
  in which each of the  sides of the simulation box were allowed to fluctuate independently.
  The number of replicas, $P$ in the path integral simulations was selected for each temperature by
  imposing that $PT$ be approximately $1900 \pm 300$.
  This choice guarantees that, for a rigid model of water, the  thermodynamic properties
  are  within two per cent of the value obtained as  $P$ tends to infinity.
  In general the simulations of this work consisted of 200,000 Monte Carlo cycles, where a cycle
  consists of a trial move per particle (the number of particles is equal to $NP$ where
  $N$ is the number of water molecules) plus a trial volume change in the case of $NpT$ simulations.
  To increase the accuracy, in the determination of the excess quantum free energies, four independent runs were performed
  for each value of $\lambda'$. In Eq. \ref{eq_guay} $\Delta \mu^{\mathrm{classical}}$ is zero if evaluated
  at the coexistence $T$ and $p$ of the classical system.
  Direct coexistence simulations were significantly longer (up to 10 million cycles)
  and Gibbs Duhem simulations were typically ten times shorter.
  Further details of the path integral Monte Carlo simulations, for example,  the rotational propagator,
  the acceptance criteria within the Markov chain, the evaluation of relative Euler angles between
  contiguous beads, can be found in \cite{MP_2011_109_0149,JCP_2011_134_054117}.
\section{RESULTS AND DISCUSSION}
 To illustrate the methodology used in this work to determine the entire phase diagram
of the quantum system, we shall describe in detail the procedure used to determine
the fluid-I$_{\mathrm {h}}$ coexistence curve:
\begin{itemize}
\item{Determine the melting temperature $T_m^{{\mathrm{classical}}}$ of the classical system at normal pressure.}
\item {Determine the integral in Eq. \ref{eq_guay} by performing path integral  $NpT$ simulations 
for various values of $\lambda'$ for ice I$_{\mathrm {h}}$ and for liquid water. For H$_2$O 
the values of $\lambda'=1,6/7,5/7,4/7,3/7,2/7,1/7$ were used to evaluate this integral.} 
\item{ Perform thermodynamic integration and determine the melting temperature  $T_m$
 of the quantum system at which ice I$_{\mathrm {h}}$ and water have the same $\mu$ at normal $p$.}
\item{ Perform Gibbs-Duhem integration using path integral simulations to determine the full I$_{\mathrm {h}}$-water coexistence
 line. }
\end{itemize}
 Free energy calculations for TIP4PQ/2005 yielded $T_m^{{\mathrm{classical}}} = 282$K for 
I$_{\mathrm {h}}$ ($p=1$ bar). The same result 282$\pm 3$K was
obtained from direct coexistence simulations.
In direct coexistence runs  \cite{JPCM_2008_20_153101,JCP_2007_126_194502}   half of the simulation box is filled
with ice and the other half with the liquid. 
$NpT$ simulations at normal pressure are performed for several temperatures. 
For temperatures above the melting point the ice within the system melts, and for temperatures below the melting point the ice phase is seen to grow.

We then proceeded to calculate the integral in   Eq. \ref{eq_guay} at normal $p$ and $T=282K$ (where the
two phases have the same chemical potential in the classical system). 
To do this the mass of the atoms in the TIP4PQ/2005 ``molecule" were incrementally increased by a factor of
$\lambda$. Such a scaling only modifies the total mass of the molecule M and the eigenvalues of the inertia tensor (i.e. the principal moments of inertia),
but leaves the geometry of the model and the location of the centre of mass unchanged. 
Seven scaling factors between $\lambda=1$ and $\lambda=7$ were used, and the corresponding exact rotational propagator at $T=282$K was calculated for each.
Beyond $\lambda=7$ the calculation of the propagator becomes prohibitively expensive to calculate, and a large number of simulations
would be required to reduce the errors to an acceptable level.
Path integral simulations were then performed, and the integrand of Eq. \ref{eq_guay} is determined 
(see Figure 1a). The integral 
over the curve formed by these results provides the difference in excess quantum Gibbs energy between 
ice I$_{\mathrm {h}}$ and water, and therefore the difference in free energies between the two phases in the quantum
system (for $T=282$K and $p=1$ bar the two phases have the same chemical potential in the classical system).
As can be seen in Figure 1a the integrand for the liquid-I$_{\mathrm {h}}$ calculation is reasonably smooth and forms an almost horizontal line. For this reason it seems reasonable to extrapolate
the integrand for  $\lambda'<1/7$ from the values obtained for larger $\lambda'$.

It can be seen that the kinetic energy of ice I$_{\mathrm {h}}$ is higher than that of water at 282K and 1bar,
indicating that nuclear quantum effects are  significantly larger
in the ice I$_{\mathrm {h}}$  phase \cite{JCP_2005_123_144506}.
For rigid models nuclear quantum 
effects are related to the strength of the intermolecular interactions which, in the case
of water, is dominated by  hydrogen bonds. In ice I$_{\mathrm {h}}$ each molecule forms four 
hydrogen bonds with the first nearest neighbours, whereas in the liquid phase this number is 
somewhat smaller. The more ``localised" character of the molecular libration in ice I$_{\mathrm {h}}$ with 
respect to the liquid leads to the higher kinetic energies observed. 
From the results shown in Figure 1a it follows that $\mu$ of 
ice I$_{\mathrm {h}}$ in the quantum
system is $0.18kT$ higher than that of water at 282K and 1bar, indicating that
the melting point of ice I$_{\mathrm {h}}$ in the quantum system is lower than that of the classical system. 
By using thermodynamic integration at constant $p$ 
we found that at 258K the chemical potential of the solid and fluid phases become identical.
This value is the $T_m$ of the quantum system. 
In order to corroborate this result, direct coexistence runs of the quantum system 
were  undertaken.  For the  runs performed at $T=266$K and  $T=262$K
the total energy increases with 
time and reaches a plateau indicating the complete melting of the ice slab. At $T=240$K we 
saw a slow growth of the ice phase (See \ref{PI_direct_melt}). At $T=252$K the energy was approximately constant along 
the run. 
This indicates that the melting point lies between $T=252$K  and $T=262$K, thus we shall adopt the intermediate value of  257K ($\pm 5$K) 
in agreement with the free energy result of 258K.

The entire I$_{\mathrm {h}}$-water coexistence line is obtained by using the Gibbs-Duhem integration method.
The Gibbs-Duhem integration method consists of a
numerical integration of the Clapeyron equation and requires simply the knowledge of the
enthalpy and volume difference between the two coexisting phases. The enthalpy of each phase
is obtained from  $H = K + U + pV$. 
Note that for each new temperature in the Gibbs-Duhem integration a propagator matrix must be
calculated as the propagator depends on the value of $PT$. 
It can be seen that the I$_{\mathrm {h}}$-water curve (\ref{fig_quant_phase_diag}) is essentially parallel to the experimental curve, shifted by $\approx 15$K
due to the lower melting point of the TIP4PQ/2005 model.
The aforementioned methodology for I$_{\mathrm {h}}$-water was applied to the remaining phase equilibria,
leading to the complete phase diagram. 
A plot of the integrand of Eq. \ref{eq_guay} with respect to $\lambda'$ is shown in Figure 1a. 
The integral of these functions leads to the difference in excess quantum free energy between the two
considered phases in units of $NkT$.
As  can be seen the integrand is rather smooth, and in most of the cases it
can be well described by a straight line.
It is worth noting that if this were always the case then one could obtain a reasonable estimate of the integral
simply by obtaining the value of $\Delta K/(\lambda'NkT)$ at $\lambda'=1/2$.
In \ref{fig_quant_phase_diag} the phase diagram of water as obtained from quantum simulations of the TIP4PQ/2005 model
of water is presented and compared to the
experimental phase diagram.
One  can see that the diagram is qualitatively correct, each 
phase is situated in the correct relation to the other phases. Furthermore, 
the gradients of the coexistence curves are also acceptable in comparison to experimental results.
The most notable discrepancy is an overall shift of 15-20K in the diagram to lower temperatures.
In \ref{fig_vol} the changes in volume along phase transitions obtained from the simulations are presented. 
It can be seen that they compare favourably with the experimental results obtained by   Bridgman in 1912 \cite{PAAAS_1912_XLVII_13_441} (\ref{fig_vol}). 
Making use of a recent publication by Loerting et al. \cite{JPC_B_2011_115_14167} where  the densities of ices I$_{\mathrm {h}}$, II, III and V in the range 77-87K  
at normal pressure were determined using a methodology known as cryoflotation, we decided to study the densities at the intermediate temperature of 82K. 
Additionally we also
considered a couple of states for the liquid, and one for ice VI at room temperature. 
The results are summarised in  \ref{tbl_density}. In general the simulation and experimental results coincide to within 1\%.
Classical simulations of TIP4P/2005 tend to overestimate the experimental densities (at 82K) 
by more than 3\% \cite{JCP_2009_131_024506}. 
Thus a quantum treatment is absolutely essential if one wishes  to describe experimental
results at low temperatures.
The results of this table can also be used to estimate the value
of the transition pressures at zero Kelvin as shown by Whalley \cite{JCP_1984_81_04087,JCP_2007_127_154518}.
The estimates obtained in this way were consistent 
with those obtained from extrapolations to zero
Kelvin of the coexistence lines. The maximum deviation found between 
the two methodologies was $\approx$700 bar, which is reasonable taking into account the 
combined uncertainty of all the calculations. 

In order to highlight the differences between quantum and classical results for the phase diagram,
the quantum and classical phase diagram of the TIP4PQ/2005
model are superimposed  in \ref{fig_classical_diag}. 
Although the diagrams are qualitatively similar there are certain features of interest that can be observed.
In the classical phase diagram the melting lines are shifted to higher temperatures, and solid-solid transitions are shifted to higher pressures
(for a given temperature) with respect to the quantum phase diagram.
Another important difference between the classical and quantum phase diagrams is that the 
region of the phase diagram occupied by ice 
II is significantly  reduced in the classical treatment. In fact in the classical system 
the ice II-III transition is shifted to much lower temperatures and 
ice II is stable only for temperatures below 80K. 
This shrinking of the ice II phase
is consistent with recent findings by Habershon and Manolopoulos \cite{PCCP_2011_13_19714}
who found that in classical simulations of the q-TIP4P/F model \cite{JCP_2009_131_024501} 
ice III occupies the region of stability of ice II. 
This indicates that nuclear quantum effects
play a significant role in determining the region of stability of ice II in the phase diagram of water.

 It would be useful to have a rational guide to understand the changes in the phase
diagram observed when including nuclear quantum effects.
For this purpose the integrand of Eq. \ref{calculating_G_exc},  which facilitates the determination 
of the quantum excess free energy,  is shown in Figure 1b at a temperature of 200K. 
The average kinetic energy of a harmonic oscillator of mass
$M=M_0/\lambda'$ and frequency $\nu=\sqrt{\lambda'} \nu_0$ is given by  \cite{JCP_2010_133_144511}:
\begin{equation}
\langle K \rangle =\frac{h\nu_0}{4} \sqrt{\lambda'} \coth \left(\frac{h\nu_0}{2k_BT} \sqrt{\lambda'} \right).
\end{equation}
Upon performing a Taylor series expansion about $\lambda'=0$ one obtains:
\begin{equation}
\frac{ \langle K \rangle -\frac{1}{2}k_BT}{\lambda'}=\frac{1}{24}\frac{(h\nu_0)^{2}}{k_BT}-\frac{1}{1440}\frac{(h\nu_0)^{4}}{(k_BT)^3}\lambda'+\mathcal{O} (\lambda')^2
\end{equation}
For the rigid water model used in this work, one can describe the solid phases by a set of
$6N$ oscillators (i.e. phonons). By assuming a unique frequency,  as in an Einstein like model, one arrives at:
\begin{equation}
\frac{ \langle K \rangle/Nk_BT  -3}{\lambda'}=\frac{1}{4} \left( \frac{h\nu_0}{k_BT} \right)^2 -\frac{1}{240} \left( \frac{h\nu_0}{k_BT} \right)^4 \lambda'+\mathcal{O} (\lambda')^2
\end{equation}
Thus for the Einstein model 
the integrand of Eq. 4  is well behaved and has both a finite value and a finite negative slope at
$\lambda'=0$ . The results presented in Figure 1b are indeed consistent with this predicted behaviour. 
From Figure 1b is is possible to estimate  $\nu_0$ from either the slope or the intercept, obtaining 
``Einstein" like frequencies between  550 cm$^{-1}$ and 450 cm$^{-1}$ depending on the phase.
These are typical values for intermolecular librations in water, which are located between 50 cm$^{-1}$ and 800 cm$^{-1}$.
A recent study has shown that one can reproduce the heat capacity of ice I$_{\mathrm {h}}$ using a selection of six fundamental frequencies 
selected from this range \cite{IJThermo_2010_13_0051}. It is worth mentioning that the TIP4PQ/2005 model also does a good job of calculating $C_p$
when used in path integral simulations \cite{JCP_2010_132_046101}. 
The excess quantum free energies are obtained from the
integration of the results of Figure 1b.
It is evident that at 200K nuclear quantum effects significantly influence the free energies
of the solid phases of water.
By comparing the results of ice II at both 1 and 4112 bar it is
seen that pressure increases the magnitude of nuclear quantum effects at a given temperature
although the increase is small ($0.08$ in $NkT$ units for the considered pressures).   
The relative ordering in which the excess free energy increases is II $\simeq$VI, V, III and finally I$_{\mathrm{h}}$.

The tetrahedral order 
parameter, $q_t$ \cite{MP_1998_93_0511,N_2001_409_00318},
was designed to measure the degree of tetrahedral ordering in liquid water:
$q_{t}$ is defined as:
\begin{equation}
\label{ec1}
q_{t}= \left\langle 1-\frac{3}{8}\sum^{3}_{j=1}\sum^{4}_{k=j+1} \left( \cos( \theta_{j,i,k}) +\frac{1}{3} \right)^2 \right\rangle 
\end{equation}
where the sum is over the four nearest (oxygen) neighbours of the oxygen of the $i$-th
water molecule. The angle $\theta_{j,i,k}$ is the angle formed by the oxygens of molecules
$j$, $i$  and $k$, oxygen $i$ forming the vertex of the angle. The tetrahedral order parameter has a value of 1 for a perfect tetrahedral network, 
and 0 for an ``ideal gas'' of oxygen centres.
We shall make use of this descriptor in order to try to rationalise our results.
The value of $q_{t}$ of each solid at T=200K was obtained by annealing the solid structure while keeping the equilibrium unit cell of the system.
In \ref{q_t_vs_g}, the excess free energy
is plotted as a function of  $q_t$ for the proton disordered ices, namely I$_{\mathrm{h}}$, III, V and
VI at a pressure of around 3600 bar and a strong correlation is evident. 
Ice II has a large value,   $q_t=0.83$, however, 
the impact of nuclear quantum effects on 
this ice phase are smaller than in the rest of the ices. 
Since ice II is the only proton ordered solid considered in this work, it is clear that in this
case the fixed relative orientations between molecules are playing an important role in determining 
the magnitude of the nuclear quantum effects. 
It would be useful to evaluate the impact of nuclear quantum effects on the fluid phase. At 200K the
fluid is highly supercooled thus is difficult to evaluate its $G^{ex,Q}$. 
The difference between  the total kinetic energy of a phase and that of the corresponding classical system
under the same conditions also provides  an estimate of the magnitude of nuclear quantum effects.
One can see in  \ref{tbl_density} that at 300K and 15400 bar the kinetic energy of  liquid water is only 
slightly lower than that of ice VI. 
This indicates that the 
magnitude  of nuclear quantum effects in the liquid is smaller than that of the ice with 
smallest  nuclear quantum effects, ice VI. This is consistent with the low value $q_{t}=0.58$  of the
tetrahedral order parameter found for the fluid phase at 300K and 15400 bar \cite{JPC_B_2011_115_06935}. 
It appears  that the importance of
nuclear quantum effects increases as the strength of the intermolecular hydrogen bonding increases.
The strength of the hydrogen bonding seems to correlate  (with the exception of ice II) with the value of the
tetrahedral order parameter. 
For example, in ice I$_{\mathrm {h}}$ the first four nearest neighbours of a given molecule are located in a perfect tetrahedral
arrangement which is the optimum situation to have a strong hydrogen bond. For the rest of the
ices (and for water) the four nearest neighbours of a molecule form a distorted tetrahedron and
therefore the strength of the hydrogen bond should decrease.  The greater the strength of the
hydrogen bond the higher the frequency associated with the librational mode, and therefore the
higher the impact of nuclear quantum effects. 

  The  change of the coexistence pressure for a certain temperature due to the inclusion 
of nuclear quantum effects can be approximated reasonably well by the expression
\begin{equation}
p-p_{\mathrm{classical}} \simeq  \frac{G^{ex,Q}_{B} - G^{ex,Q}_{A}}{V_{B} - V_{A}} 
\label{impact}
\end{equation}
where the properties on the right hand side are evaluated at $p_{\mathrm{classical}}$. 
It follows from Eq. \ref{impact} that 
the impact of nuclear quantum effects on a given phase transition depends on the 
difference of the excess quantum free energy between the two phases, and on the volume
change. The excess free energy difference 
between phases  decreases in the following order: 
liquid-I$_{\mathrm{h}}$, II-I$_{\mathrm{h}}$, liquid-III, II-III, liquid-V,  III-I$_{\mathrm{h}}$, III-V, liquid-VI and  finally II-VI.
The impact of nuclear quantum effects on a certain phase transition will be small when 
the volume change of the phase transition is large, and large when the volume change is 
small. Volume change along the phase transitions
of water are presented in \ref{fig_vol}. Taking these two factors into account 
it is clear that the II-III phase transition is most affected by nuclear quantum 
effects (i.e. the excess free energy difference is large and the volume change is small), 
followed by the melting curves of ices (decreasing in the order liquid-I$_{\mathrm{h}}$, liquid-III, liquid-V).
This is followed by the transitions  I$_{\mathrm{h}}$-II, 
III-V and I$_{\mathrm{h}}$-III. Finally the liquid-VI and the II-VI coexistence lines are
those least  affected  by nuclear quantum effects.

\section{CONCLUSION}
This work illustrates that the calculation of the phase diagram of water,  including nuclear quantum effects, is now feasible, although admittedly it is computationally expensive  
(even for the simple model considered in this work
8 CPU's were required for about 2 years to obtain the  phase diagram presented). 
The impact of nuclear quantum effects on phase transitions is significant
and can be rationalised in terms of the degree of tetrahedral ordering of the
different phases and of the magnitude of the volume change involved in each phase transition. 
The TIP4PQ/2005 model  yields a  reasonable prediction of the experimental 
phase diagram of water. 
The simulation results are consistent with the Third Law of thermodynamics and
predict rather well the densities of
the different phases over a wide range of temperatures and pressures. 
With some delay with respect to the original contributors, 
this work shows that a simple modification of the water model proposed by
Bernal and Fowler \cite{JCP_1933_01_00515} in 1933, can
reproduce reasonably well the experimental phase diagram of water determined by Bridgmann \cite{PAAAS_1912_XLVII_13_441} in 1912, providing 
results at low temperatures consistent with the Third Law first stated by Nernst in 1906. 

In concluding this work it is worth commenting on the use of a rigid non-polarisable model to represent water.
Naturally in reality  water is both flexible and polarisable  \cite{PRL_2011_107_185701}
so it goes without saying that this work is far from the last word on the matter, and the results presented here
form only a way-point on the long road to obtaining a definitive model of water that describes all
of the facets of this intriguing molecule.
That said, path integral simulations of the TIP4PQ/2005 model has provided us with the best phase diagram of water calculated to-date.

This work was funded by grants FIS2010-16159 and FIS2010-15502 of the  Direcci\'on General de Investigaci\'on
and  S2009/ESP-1691-QF-UCM (MODELICO) of the Comunidad Aut\'onoma de Madrid.
The authors would like to thank Prof. J. L. F. Abascal for many helpful discussions.

\begin{figure}
\centerline{\includegraphics[angle=0,width=87mm,clip]{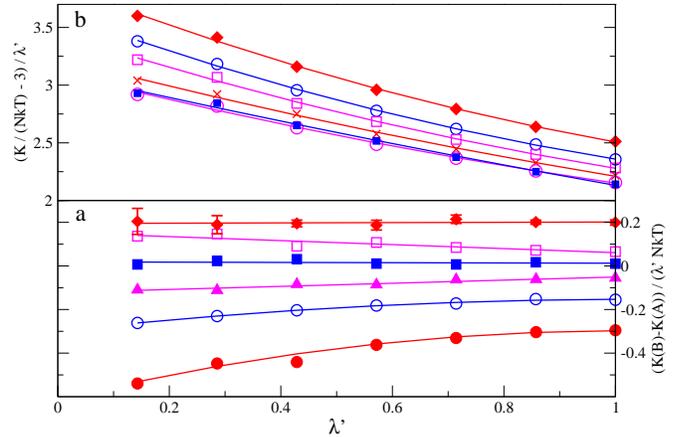}}
\caption{
\label{fig_integrand_282} 
(a) Integrand of Eq. \ref{eq_guay}
 (i.e. $(K_{B}-K_{A})/(\lambda'NkT)$) as a function of $\lambda'$ for transitions $A-B$.
Key:  red line with $\blacklozenge$ is liquid-I$_{\mathrm {h}}$ at 282K and $p=1$bar,
magenta line with $\square$ is II-V at 200K and $p=4112$bar,
blue line with $\blacksquare$ is II-VI at 200K and $p=1$bar,
magenta line with $\blacktriangle$ is V-VI at 200K and $p=9505$bar,
blue line with $\bigcirc$ is I$_{\mathrm {h}}$-III at 200K and $p=3306$bar,
and the red line with $\bullet$ is I$_{\mathrm {h}}$-II at 200K and $p=1$bar.
Error bars (only shown for liquid-I$_{\mathrm {h}}$) represent the standard error.
(b)  Integrand of Eq. \ref{calculating_G_exc} for several ices at
200K. Results were obtained using path integral simulations of the TIP4PQ/2005 model. The lines correspond
to a fit of the simulation results to a second order polynomial. Results (from top to
bottom) correspond to ice I$_{\mathrm{h}}$ at 3306bar, ice III at 3306bar, ice V at 4112bar, ice II at 4112bar,
ice II at 1bar and ice VI at 1 bar.
The integral of the curves (from 0 to 1) yields $G^{ex,Q}/(NkT)$ which  results in
(from top to bottom) 3.11, 2.91, 2.80, 2.68, 2.60 and 2.59.}
\end{figure}
\begin{figure}[t!]
\begin{center}
\includegraphics[width=87mm,angle=0,clip]{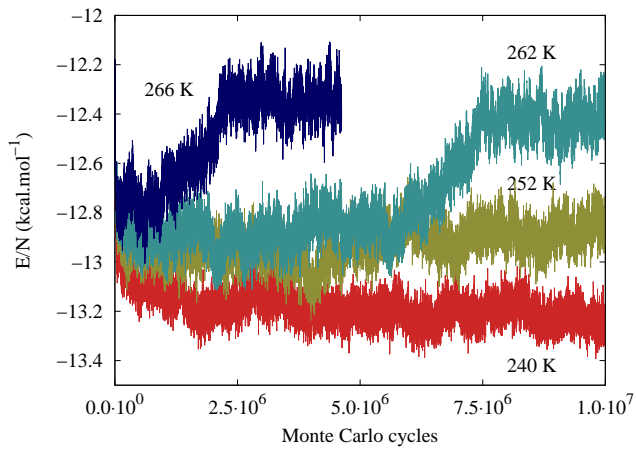}
\caption{\label{PI_direct_melt}
Plot of the total energy ($E$) per particle from the liquid-I$_{\mathrm {h}}$ direct coexistence simulations of the quantum system at $p=1$ bar.
}
\end{center}
\end{figure}
\begin{figure}[t]
\begin{center}
\includegraphics[clip,angle=270,width=87mm,clip]{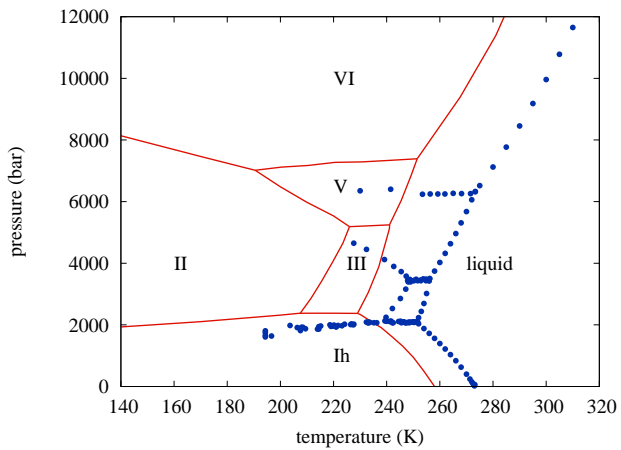}
\caption{\label{fig_quant_phase_diag} Phase diagram of water from path integral  simulations of the TIP4PQ/2005 model.
Experimental results (blue points) are also shown \cite{PAAAS_1912_XLVII_13_441,JCP_1980_73_02454}
}
\end{center}
\end{figure}
\begin{figure}[t]
\begin{center}
\includegraphics[angle=270,width=87mm,clip]{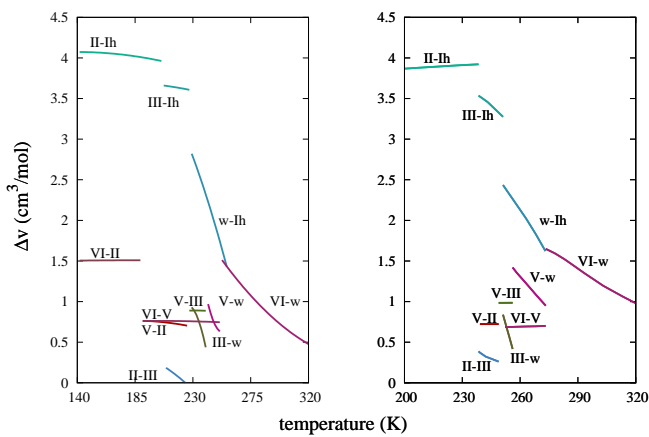}
\caption{Molar volume change ($\Delta v = v_B -v_A$) along the phase boundaries from  (Left)  path integral simulations  and (Right) experimental results \cite{PAAAS_1912_XLVII_13_441}. (w indicates the liquid phase).}
\label{fig_vol}
\end{center}
\end{figure}

\begin{figure}[t]
\begin{center}
\includegraphics[angle=270,width=87mm,clip]{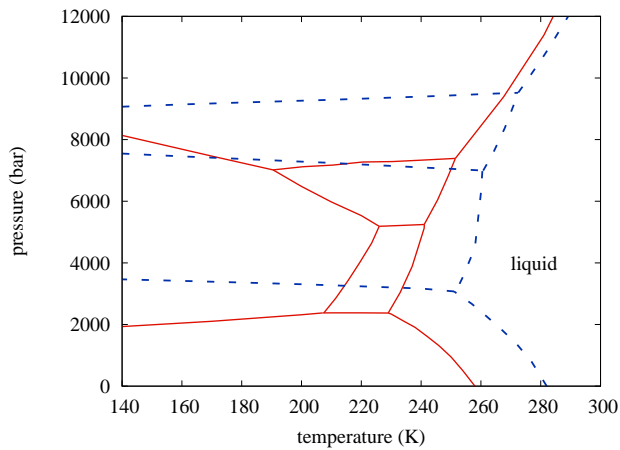}
\caption{\label{fig_classical_diag} Classical phase diagram of the TIP4PQ/2005 model (dashed blue lines)
compared to the diagram obtained from path integral simulations (solid red lines).}
\end{center}
\end{figure}
\begin{figure}[t]
\begin{center}
\includegraphics[angle=0,width=87mm,clip]{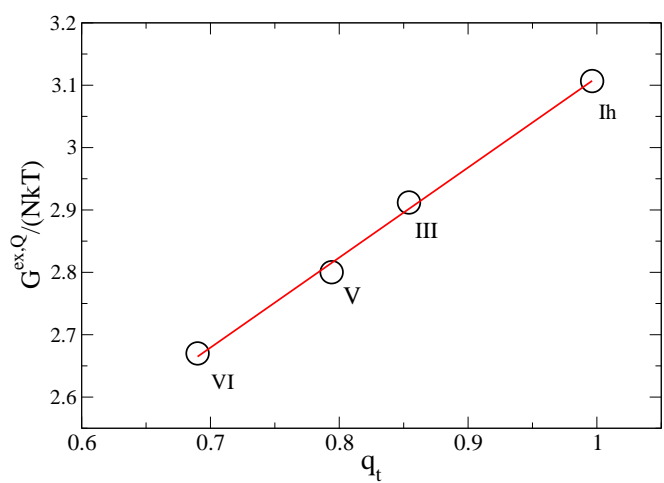}
\caption{\label{q_t_vs_g} Correlation between the excess quantum free energy and
the tetrahedral order parameter for ices at 200K and $p \approx 3600$ bar. For ice VI the excess
quantum free energy at 1 bar was increased by 0.08NkT to estimate its value at 3600 bar.}
\end{center}
\end{figure}
\clearpage
\begin{table*}[h]
\caption{\label{tbl_density}
Densities of ices and liquid of water (in g/cm$^3$) under  different thermodynamic conditions. All results were obtained
for $p=1$bar, except the two labelled with an asterisk for which $p=15400$ bar.
Experimental densities for ice III correspond to
the experimental values for ice IX ($^a$ \cite{JPC_B_2011_115_14167}, $^b$ \cite{ACSB_1968_24_1317_nolotengo}) the proton ordered form of ice III.
Experimental results are from Refs.  \cite{JPC_B_2011_115_14167,JAC_2005_38_0612,JPC_B_2011_115_14167,JPCRD_1989_18_1537,ISIS_RB_1010211}
The average values of $U$ and $K$ obtained from simulations (kcal/mol) are also shown. }
\begin{tabular}{lrcccccc}
Phase   &  $T$(K)  &     $\rho$  (sim.) &   $\rho$ (exp.) &  $U$ &   $K$     \\
\hline
I$_{\mathrm {h}}$ & 82                    & 0.927        &  0.932                  & -14.302  &   1.914      \\   
II  &               82                    & 1.189        &  1.211                  & -14.136 &   1.774  \\   

II  &               123                   & 1.185        &  1.190                  & -14.046 &   1.837   \\   

III &               82                    & 1.148        &  1.169$^a$,1.160$^b$                 & -14.040  &   1.863  \\   
V   &               82                    & 1.252        &  1.249                  & -13.883 &   1.808  \\   
VI  &               82                    & 1.335        &  1.335                  & -13.745 &   1.790  \\   
VI(*)      &           300          & 1.383        &  1.391                  & -13.055 &   2.475 \\   
liquid(*)  &           300          & 1.312        &  1.311                  & -11.997 &   2.395  \\   
liquid  &          300                   & 0.997        &  0.996                 & -11.897 &   2.366    \\   
\hline
\end{tabular}
\end{table*}

\footnotesize{
\bibliography{./bibliography} 

\providecommand*{\mcitethebibliography}{\thebibliography}
\csname @ifundefined\endcsname{endmcitethebibliography}
{\let\endmcitethebibliography\endthebibliography}{}
\begin{mcitethebibliography}{60}
\providecommand*{\natexlab}[1]{#1}
\providecommand*{\mciteSetBstSublistMode}[1]{}
\providecommand*{\mciteSetBstMaxWidthForm}[2]{}
\providecommand*{\mciteBstWouldAddEndPuncttrue}
  {\def\EndOfBibitem{\unskip.}}
\providecommand*{\mciteBstWouldAddEndPunctfalse}
  {\let\EndOfBibitem\relax}
\providecommand*{\mciteSetBstMidEndSepPunct}[3]{}
\providecommand*{\mciteSetBstSublistLabelBeginEnd}[3]{}
\providecommand*{\EndOfBibitem}{}
\mciteSetBstSublistMode{f}
\mciteSetBstMaxWidthForm{subitem}
{(\emph{\alph{mcitesubitemcount}})}
\mciteSetBstSublistLabelBeginEnd{\mcitemaxwidthsubitemform\space}
{\relax}{\relax}

\bibitem[Bridgman(1912)]{PAAAS_1912_XLVII_13_441}
P.~W. Bridgman, \emph{Proc. Amer. Acad. Arts Sci.}, 1912, \textbf{XLVII},
  441\relax
\mciteBstWouldAddEndPuncttrue
\mciteSetBstMidEndSepPunct{\mcitedefaultmidpunct}
{\mcitedefaultendpunct}{\mcitedefaultseppunct}\relax
\EndOfBibitem
\bibitem[Salzmann \emph{et~al.}(2009)Salzmann, Radaelli, Mayer, and
  Finney]{PRL_2009_103_105701}
C.~G. Salzmann, P.~G. Radaelli, E.~Mayer and J.~L. Finney, \emph{Phys. Rev.
  Lett.}, 2009, \textbf{103}, 105701\relax
\mciteBstWouldAddEndPuncttrue
\mciteSetBstMidEndSepPunct{\mcitedefaultmidpunct}
{\mcitedefaultendpunct}{\mcitedefaultseppunct}\relax
\EndOfBibitem
\bibitem[Salzmann \emph{et~al.}(2011)Salzmann, Radaelli, Slater, and
  Finney]{PCCP_2011_13_18468}
C.~G. Salzmann, P.~G. Radaelli, B.~Slater and J.~L. Finney, \emph{Phys. Chem.
  Chem. Phys.}, 2011, \textbf{13}, 18468\relax
\mciteBstWouldAddEndPuncttrue
\mciteSetBstMidEndSepPunct{\mcitedefaultmidpunct}
{\mcitedefaultendpunct}{\mcitedefaultseppunct}\relax
\EndOfBibitem
\bibitem[Morse and Rice(1982)]{JCP_1982_76_00650}
M.~D. Morse and S.~A. Rice, \emph{J. Chem. Phys.}, 1982, \textbf{76}, 650\relax
\mciteBstWouldAddEndPuncttrue
\mciteSetBstMidEndSepPunct{\mcitedefaultmidpunct}
{\mcitedefaultendpunct}{\mcitedefaultseppunct}\relax
\EndOfBibitem
\bibitem[Whalley(1984)]{JCP_1984_81_04087}
E.~Whalley, \emph{J. Chem. Phys.}, 1984, \textbf{81}, 4087\relax
\mciteBstWouldAddEndPuncttrue
\mciteSetBstMidEndSepPunct{\mcitedefaultmidpunct}
{\mcitedefaultendpunct}{\mcitedefaultseppunct}\relax
\EndOfBibitem
\bibitem[Barker and Watts(1969)]{CPL_1969_03_0144}
J.~A. Barker and R.~O. Watts, \emph{Chem. Phys. Lett.}, 1969, \textbf{3},
  144\relax
\mciteBstWouldAddEndPuncttrue
\mciteSetBstMidEndSepPunct{\mcitedefaultmidpunct}
{\mcitedefaultendpunct}{\mcitedefaultseppunct}\relax
\EndOfBibitem
\bibitem[Rahman and Stillinger(1971)]{JCP_1971_55_03336}
A.~Rahman and F.~H. Stillinger, \emph{J. Chem. Phys.}, 1971, \textbf{55},
  3336\relax
\mciteBstWouldAddEndPuncttrue
\mciteSetBstMidEndSepPunct{\mcitedefaultmidpunct}
{\mcitedefaultendpunct}{\mcitedefaultseppunct}\relax
\EndOfBibitem
\bibitem[Sanz \emph{et~al.}(2004)Sanz, Vega, Abascal, and
  MacDowell]{PRL_2004_92_255701}
E.~Sanz, C.~Vega, J.~L.~F. Abascal and L.~G. MacDowell, \emph{Phys. Rev.
  Lett.}, 2004, \textbf{92}, 255701\relax
\mciteBstWouldAddEndPuncttrue
\mciteSetBstMidEndSepPunct{\mcitedefaultmidpunct}
{\mcitedefaultendpunct}{\mcitedefaultseppunct}\relax
\EndOfBibitem
\bibitem[Abascal and Vega(2005)]{JCP_2005_123_234505}
J.~L.~F. Abascal and C.~Vega, \emph{J. Chem. Phys.}, 2005, \textbf{123},
  234505\relax
\mciteBstWouldAddEndPuncttrue
\mciteSetBstMidEndSepPunct{\mcitedefaultmidpunct}
{\mcitedefaultendpunct}{\mcitedefaultseppunct}\relax
\EndOfBibitem
\bibitem[Vega and Abascal(2011)]{PCCP_2011_13_19663}
C.~Vega and J.~L.~F. Abascal, \emph{Phys. Chem. Chem. Phys.}, 2011,
  \textbf{13}, 19663\relax
\mciteBstWouldAddEndPuncttrue
\mciteSetBstMidEndSepPunct{\mcitedefaultmidpunct}
{\mcitedefaultendpunct}{\mcitedefaultseppunct}\relax
\EndOfBibitem
\bibitem[Aragones \emph{et~al.}(2011)Aragones, MacDowell, Siepmann, and
  Vega]{PRL_2011_107_155702}
J.~L. Aragones, L.~G. MacDowell, J.~I. Siepmann and C.~Vega, \emph{Phys. Rev.
  Lett.}, 2011, \textbf{107}, 155702\relax
\mciteBstWouldAddEndPuncttrue
\mciteSetBstMidEndSepPunct{\mcitedefaultmidpunct}
{\mcitedefaultendpunct}{\mcitedefaultseppunct}\relax
\EndOfBibitem
\bibitem[McBride \emph{et~al.}(2009)McBride, Vega, Noya, Ram{\'{\i}}rez, and
  Ses{\'e}]{JCP_2009_131_024506}
C.~McBride, C.~Vega, E.~G. Noya, R.~Ram{\'{\i}}rez and L.~M. Ses{\'e}, \emph{J.
  Chem. Phys.}, 2009, \textbf{131}, 024506\relax
\mciteBstWouldAddEndPuncttrue
\mciteSetBstMidEndSepPunct{\mcitedefaultmidpunct}
{\mcitedefaultendpunct}{\mcitedefaultseppunct}\relax
\EndOfBibitem
\bibitem[Vega \emph{et~al.}(2010)Vega, Conde, McBride, Abascal, Noya,
  Ram{\'{\i}}rez, and Ses{\'e}]{JCP_2010_132_046101}
C.~Vega, M.~M. Conde, C.~McBride, J.~L.~F. Abascal, E.~G. Noya,
  R.~Ram{\'{\i}}rez and L.~M. Ses{\'e}, \emph{J. Chem. Phys.}, 2010,
  \textbf{132}, 046101\relax
\mciteBstWouldAddEndPuncttrue
\mciteSetBstMidEndSepPunct{\mcitedefaultmidpunct}
{\mcitedefaultendpunct}{\mcitedefaultseppunct}\relax
\EndOfBibitem
\bibitem[de~la Pe\~{n}a and Kusalik(2005)]{JACS_2005_127_05246}
L.~H. de~la Pe\~{n}a and P.~G. Kusalik, \emph{J. Am. Chem. Soc.}, 2005,
  \textbf{127}, 5246\relax
\mciteBstWouldAddEndPuncttrue
\mciteSetBstMidEndSepPunct{\mcitedefaultmidpunct}
{\mcitedefaultendpunct}{\mcitedefaultseppunct}\relax
\EndOfBibitem
\bibitem[Fanourgakis and Xantheas(2008)]{JCP_2008_128_074506}
G.~S. Fanourgakis and S.~S. Xantheas, \emph{J. Chem. Phys.}, 2008,
  \textbf{128}, 074506\relax
\mciteBstWouldAddEndPuncttrue
\mciteSetBstMidEndSepPunct{\mcitedefaultmidpunct}
{\mcitedefaultendpunct}{\mcitedefaultseppunct}\relax
\EndOfBibitem
\bibitem[Morrone and Car(2008)]{PRL_2008_101_017801}
J.~A. Morrone and R.~Car, \emph{Phys. Rev. Lett.}, 2008, \textbf{101},
  017801\relax
\mciteBstWouldAddEndPuncttrue
\mciteSetBstMidEndSepPunct{\mcitedefaultmidpunct}
{\mcitedefaultendpunct}{\mcitedefaultseppunct}\relax
\EndOfBibitem
\bibitem[Paesani and Voth(2009)]{JPCB_2009_113_05702}
F.~Paesani and G.~A. Voth, \emph{J. Phys. Chem. B}, 2009, \textbf{113},
  5702\relax
\mciteBstWouldAddEndPuncttrue
\mciteSetBstMidEndSepPunct{\mcitedefaultmidpunct}
{\mcitedefaultendpunct}{\mcitedefaultseppunct}\relax
\EndOfBibitem
\bibitem[Ceriotti \emph{et~al.}(2010)Ceriotti, Parrinello, Markland, and
  Manolopoulos]{JCP_2010_133_124104}
M.~Ceriotti, M.~Parrinello, T.~E. Markland and D.~E. Manolopoulos, \emph{J.
  Chem. Phys.}, 2010, \textbf{133}, 124104\relax
\mciteBstWouldAddEndPuncttrue
\mciteSetBstMidEndSepPunct{\mcitedefaultmidpunct}
{\mcitedefaultendpunct}{\mcitedefaultseppunct}\relax
\EndOfBibitem
\bibitem[Habershon and Manolopoulos(2011)]{JCP_2011_135_224111}
S.~Habershon and D.~E. Manolopoulos, \emph{J. Chem. Phys.}, 2011, \textbf{135},
  224111\relax
\mciteBstWouldAddEndPuncttrue
\mciteSetBstMidEndSepPunct{\mcitedefaultmidpunct}
{\mcitedefaultendpunct}{\mcitedefaultseppunct}\relax
\EndOfBibitem
\bibitem[Feynman(1948)]{RMP_1948_20_00367}
R.~P. Feynman, \emph{Rev. Modern Phys.}, 1948, \textbf{20}, 367\relax
\mciteBstWouldAddEndPuncttrue
\mciteSetBstMidEndSepPunct{\mcitedefaultmidpunct}
{\mcitedefaultendpunct}{\mcitedefaultseppunct}\relax
\EndOfBibitem
\bibitem[Gillan(1990)]{NATO_ASI_C_293_0155_photocopy}
M.~J. Gillan, in \emph{The path-integral simulation of quantum systems}, ed.
  C.~R.~A. Catlow, S.~C. Parker and M.~P. Allen, Kluwer, The Netherlands, 1990,
  vol. 293, ch.~6, pp. 155--188\relax
\mciteBstWouldAddEndPuncttrue
\mciteSetBstMidEndSepPunct{\mcitedefaultmidpunct}
{\mcitedefaultendpunct}{\mcitedefaultseppunct}\relax
\EndOfBibitem
\bibitem[Barker(1979)]{JCP_1979_70_02914}
J.~A. Barker, \emph{J. Chem. Phys.}, 1979, \textbf{70}, 2914\relax
\mciteBstWouldAddEndPuncttrue
\mciteSetBstMidEndSepPunct{\mcitedefaultmidpunct}
{\mcitedefaultendpunct}{\mcitedefaultseppunct}\relax
\EndOfBibitem
\bibitem[Chandler and Wolynes(1981)]{JCP_1981_74_04078}
D.~Chandler and P.~G. Wolynes, \emph{J. Chem. Phys.}, 1981, \textbf{74},
  4078\relax
\mciteBstWouldAddEndPuncttrue
\mciteSetBstMidEndSepPunct{\mcitedefaultmidpunct}
{\mcitedefaultendpunct}{\mcitedefaultseppunct}\relax
\EndOfBibitem
\bibitem[Sedlmeier \emph{et~al.}(2011)Sedlmeier, Horinek, and
  Netz]{JACS_2011_133_01391}
F.~Sedlmeier, D.~Horinek and R.~R. Netz, \emph{J. Am. Chem. Soc.}, 2011,
  \textbf{133}, 1391\relax
\mciteBstWouldAddEndPuncttrue
\mciteSetBstMidEndSepPunct{\mcitedefaultmidpunct}
{\mcitedefaultendpunct}{\mcitedefaultseppunct}\relax
\EndOfBibitem
\bibitem[Sancho and Best(2011)]{JACS_2011_133_06809}
D.~D. Sancho and R.~B. Best, \emph{J. Am. Chem. Soc.}, 2011, \textbf{133},
  6809\relax
\mciteBstWouldAddEndPuncttrue
\mciteSetBstMidEndSepPunct{\mcitedefaultmidpunct}
{\mcitedefaultendpunct}{\mcitedefaultseppunct}\relax
\EndOfBibitem
\bibitem[Noya \emph{et~al.}(2009)Noya, Vega, Ses{\'e}, and
  Ram{\'{\i}}rez]{JCP_2009_131_124518}
E.~G. Noya, C.~Vega, L.~M. Ses{\'e} and R.~Ram{\'{\i}}rez, \emph{J. Chem.
  Phys.}, 2009, \textbf{131}, 124518\relax
\mciteBstWouldAddEndPuncttrue
\mciteSetBstMidEndSepPunct{\mcitedefaultmidpunct}
{\mcitedefaultendpunct}{\mcitedefaultseppunct}\relax
\EndOfBibitem
\bibitem[Wallqvist and Berne(1985)]{CPL_1985_117_0214}
A.~Wallqvist and B.~J. Berne, \emph{Chem. Phys. Lett.}, 1985, \textbf{117},
  214\relax
\mciteBstWouldAddEndPuncttrue
\mciteSetBstMidEndSepPunct{\mcitedefaultmidpunct}
{\mcitedefaultendpunct}{\mcitedefaultseppunct}\relax
\EndOfBibitem
\bibitem[Kuharski and Rossky(1984)]{CPL_1984_103_0357}
R.~A. Kuharski and P.~J. Rossky, \emph{Chem. Phys. Lett.}, 1984, \textbf{103},
  357\relax
\mciteBstWouldAddEndPuncttrue
\mciteSetBstMidEndSepPunct{\mcitedefaultmidpunct}
{\mcitedefaultendpunct}{\mcitedefaultseppunct}\relax
\EndOfBibitem
\bibitem[Curotto \emph{et~al.}(2008)Curotto, Freeman, and
  Doll]{JCP_2008_128_204107}
E.~Curotto, D.~L. Freeman and J.~D. Doll, \emph{J. Chem. Phys.}, 2008,
  \textbf{128}, 204107\relax
\mciteBstWouldAddEndPuncttrue
\mciteSetBstMidEndSepPunct{\mcitedefaultmidpunct}
{\mcitedefaultendpunct}{\mcitedefaultseppunct}\relax
\EndOfBibitem
\bibitem[Asare \emph{et~al.}(2009)Asare, Musah, Curotto, Freeman, and
  Doll]{JCP_2009_131_184508}
E.~Asare, A.-R. Musah, E.~Curotto, D.~L. Freeman and J.~D. Doll, \emph{J. Chem.
  Phys.}, 2009, \textbf{131}, 184508\relax
\mciteBstWouldAddEndPuncttrue
\mciteSetBstMidEndSepPunct{\mcitedefaultmidpunct}
{\mcitedefaultendpunct}{\mcitedefaultseppunct}\relax
\EndOfBibitem
\bibitem[M{\"u}ser and Berne(1996)]{PRL_1996_77_002638}
M.~H. M{\"u}ser and B.~J. Berne, \emph{Phys. Rev. Lett.}, 1996, \textbf{77},
  2638\relax
\mciteBstWouldAddEndPuncttrue
\mciteSetBstMidEndSepPunct{\mcitedefaultmidpunct}
{\mcitedefaultendpunct}{\mcitedefaultseppunct}\relax
\EndOfBibitem
\bibitem[Noya \emph{et~al.}(2011)Noya, Vega, and McBride]{JCP_2011_134_054117}
E.~G. Noya, C.~Vega and C.~McBride, \emph{J. Chem. Phys.}, 2011, \textbf{134},
  054117\relax
\mciteBstWouldAddEndPuncttrue
\mciteSetBstMidEndSepPunct{\mcitedefaultmidpunct}
{\mcitedefaultendpunct}{\mcitedefaultseppunct}\relax
\EndOfBibitem
\bibitem[Noya \emph{et~al.}(2011)Noya, Ses{\'e}, Ram{\'{\i}}rez, McBride,
  Conde, and Vega]{MP_2011_109_0149}
E.~G. Noya, L.~M. Ses{\'e}, R.~Ram{\'{\i}}rez, C.~McBride, M.~M. Conde and
  C.~Vega, \emph{Molec. Phys.}, 2011, \textbf{109}, 149\relax
\mciteBstWouldAddEndPuncttrue
\mciteSetBstMidEndSepPunct{\mcitedefaultmidpunct}
{\mcitedefaultendpunct}{\mcitedefaultseppunct}\relax
\EndOfBibitem
\bibitem[Frenkel and Ladd(1984)]{JCP_1984_81_03188}
D.~Frenkel and A.~J.~C. Ladd, \emph{J. Chem. Phys.}, 1984, \textbf{81},
  3188\relax
\mciteBstWouldAddEndPuncttrue
\mciteSetBstMidEndSepPunct{\mcitedefaultmidpunct}
{\mcitedefaultendpunct}{\mcitedefaultseppunct}\relax
\EndOfBibitem
\bibitem[Noya \emph{et~al.}(2008)Noya, Conde, and Vega]{JCP_2008_129_104704}
E.~G. Noya, M.~M. Conde and C.~Vega, \emph{J. Chem. Phys.}, 2008, \textbf{129},
  104704\relax
\mciteBstWouldAddEndPuncttrue
\mciteSetBstMidEndSepPunct{\mcitedefaultmidpunct}
{\mcitedefaultendpunct}{\mcitedefaultseppunct}\relax
\EndOfBibitem
\bibitem[Johnson \emph{et~al.}(1993)Johnson, Zollweg, and
  Gubbins]{MP_1993_78_0591}
J.~K. Johnson, J.~A. Zollweg and K.~E. Gubbins, \emph{Molec. Phys.}, 1993,
  \textbf{78}, 591\relax
\mciteBstWouldAddEndPuncttrue
\mciteSetBstMidEndSepPunct{\mcitedefaultmidpunct}
{\mcitedefaultendpunct}{\mcitedefaultseppunct}\relax
\EndOfBibitem
\bibitem[Kofke(1993)]{JCP_1993_98_04149}
D.~A. Kofke, \emph{J. Chem. Phys.}, 1993, \textbf{98}, 4149\relax
\mciteBstWouldAddEndPuncttrue
\mciteSetBstMidEndSepPunct{\mcitedefaultmidpunct}
{\mcitedefaultendpunct}{\mcitedefaultseppunct}\relax
\EndOfBibitem
\bibitem[Vega \emph{et~al.}(2008)Vega, Sanz, Abascal, and
  Noya]{JPCM_2008_20_153101}
C.~Vega, E.~Sanz, J.~L.~F. Abascal and E.~G. Noya, \emph{J. Phys. Cond. Mat.},
  2008, \textbf{20}, 153101\relax
\mciteBstWouldAddEndPuncttrue
\mciteSetBstMidEndSepPunct{\mcitedefaultmidpunct}
{\mcitedefaultendpunct}{\mcitedefaultseppunct}\relax
\EndOfBibitem
\bibitem[Beck(2007)]{libro_free_energy}
T.~L. Beck, in \emph{Quantum contributions to free energy changes in fluids},
  ed. C.~Chipot and A.~Pohorille, Springer-Verlag, Berling, Germany, 2007,
  vol.~86, p. 389\relax
\mciteBstWouldAddEndPuncttrue
\mciteSetBstMidEndSepPunct{\mcitedefaultmidpunct}
{\mcitedefaultendpunct}{\mcitedefaultseppunct}\relax
\EndOfBibitem
\bibitem[Buch \emph{et~al.}(1998)Buch, Sandler, and
  Sadlej]{JPCB_1998_102_08641}
V.~Buch, P.~Sandler and J.~Sadlej, \emph{J. Phys. Chem. B}, 1998, \textbf{102},
  8641\relax
\mciteBstWouldAddEndPuncttrue
\mciteSetBstMidEndSepPunct{\mcitedefaultmidpunct}
{\mcitedefaultendpunct}{\mcitedefaultseppunct}\relax
\EndOfBibitem
\bibitem[Bernal and Fowler(1933)]{JCP_1933_01_00515}
J.~D. Bernal and R.~H. Fowler, \emph{J. Chem. Phys.}, 1933, \textbf{1},
  515\relax
\mciteBstWouldAddEndPuncttrue
\mciteSetBstMidEndSepPunct{\mcitedefaultmidpunct}
{\mcitedefaultendpunct}{\mcitedefaultseppunct}\relax
\EndOfBibitem
\bibitem[Pauling(1935)]{JACS_1935_57_02680}
L.~Pauling, \emph{J. Am. Chem. Soc.}, 1935, \textbf{57}, 2680\relax
\mciteBstWouldAddEndPuncttrue
\mciteSetBstMidEndSepPunct{\mcitedefaultmidpunct}
{\mcitedefaultendpunct}{\mcitedefaultseppunct}\relax
\EndOfBibitem
\bibitem[Aragones \emph{et~al.}(2011)Aragones, MacDowell, and
  Vega]{JPC_A_2011_115_005745}
J.~L. Aragones, L.~G. MacDowell and C.~Vega, \emph{J. Phys. Chem. A}, 2011,
  \textbf{115}, 5745\relax
\mciteBstWouldAddEndPuncttrue
\mciteSetBstMidEndSepPunct{\mcitedefaultmidpunct}
{\mcitedefaultendpunct}{\mcitedefaultseppunct}\relax
\EndOfBibitem
\bibitem[Cazorla \emph{et~al.}(2007)Cazorla, Gillan, Taioli, and
  Alfe]{JCP_2007_126_194502}
C.~Cazorla, M.~J. Gillan, S.~Taioli and D.~Alfe, \emph{J. Chem. Phys.}, 2007,
  \textbf{126}, 194502\relax
\mciteBstWouldAddEndPuncttrue
\mciteSetBstMidEndSepPunct{\mcitedefaultmidpunct}
{\mcitedefaultendpunct}{\mcitedefaultseppunct}\relax
\EndOfBibitem
\bibitem[de~la Pe\~{n}a \emph{et~al.}(2005)de~la Pe\~{n}a, Razul, and
  Kusalik]{JCP_2005_123_144506}
L.~H. de~la Pe\~{n}a, M.~S.~G. Razul and P.~G. Kusalik, \emph{J. Chem. Phys.},
  2005, \textbf{123}, 144506\relax
\mciteBstWouldAddEndPuncttrue
\mciteSetBstMidEndSepPunct{\mcitedefaultmidpunct}
{\mcitedefaultendpunct}{\mcitedefaultseppunct}\relax
\EndOfBibitem
\bibitem[Loerting \emph{et~al.}(2011)Loerting, Bauer, Kohl, Watschinger,
  Winkel, and Mayer]{JPC_B_2011_115_14167}
T.~Loerting, M.~Bauer, I.~Kohl, K.~Watschinger, K.~Winkel and E.~Mayer,
  \emph{J. Phys. Chem. B}, 2011, \textbf{115}, 14167\relax
\mciteBstWouldAddEndPuncttrue
\mciteSetBstMidEndSepPunct{\mcitedefaultmidpunct}
{\mcitedefaultendpunct}{\mcitedefaultseppunct}\relax
\EndOfBibitem
\bibitem[Aragones \emph{et~al.}(2007)Aragones, Noya, Abascal, and
  Vega]{JCP_2007_127_154518}
J.~L. Aragones, E.~G. Noya, J.~L.~F. Abascal and C.~Vega, \emph{J. Chem.
  Phys.}, 2007, \textbf{127}, 154518\relax
\mciteBstWouldAddEndPuncttrue
\mciteSetBstMidEndSepPunct{\mcitedefaultmidpunct}
{\mcitedefaultendpunct}{\mcitedefaultseppunct}\relax
\EndOfBibitem
\bibitem[Habershon and Manolopoulos(2011)]{PCCP_2011_13_19714}
S.~Habershon and D.~E. Manolopoulos, \emph{Phys. Chem. Chem. Phys.}, 2011,
  \textbf{13}, 19714\relax
\mciteBstWouldAddEndPuncttrue
\mciteSetBstMidEndSepPunct{\mcitedefaultmidpunct}
{\mcitedefaultendpunct}{\mcitedefaultseppunct}\relax
\EndOfBibitem
\bibitem[Habershon \emph{et~al.}(2009)Habershon, Markland, and
  Manolopoulos]{JCP_2009_131_024501}
S.~Habershon, T.~E. Markland and D.~E. Manolopoulos, \emph{J. Chem. Phys.},
  2009, \textbf{131}, 024501\relax
\mciteBstWouldAddEndPuncttrue
\mciteSetBstMidEndSepPunct{\mcitedefaultmidpunct}
{\mcitedefaultendpunct}{\mcitedefaultseppunct}\relax
\EndOfBibitem
\bibitem[Ram{\'{\i}}rez and Herrero(2010)]{JCP_2010_133_144511}
R.~Ram{\'{\i}}rez and C.~P. Herrero, \emph{J. Chem. Phys.}, 2010, \textbf{133},
  144511\relax
\mciteBstWouldAddEndPuncttrue
\mciteSetBstMidEndSepPunct{\mcitedefaultmidpunct}
{\mcitedefaultendpunct}{\mcitedefaultseppunct}\relax
\EndOfBibitem
\bibitem[Wang and Brewster(2010)]{IJThermo_2010_13_0051}
K.-T. Wang and Q.~M. Brewster, \emph{Int. J. Thermodynamics}, 2010,
  \textbf{13}, 51\relax
\mciteBstWouldAddEndPuncttrue
\mciteSetBstMidEndSepPunct{\mcitedefaultmidpunct}
{\mcitedefaultendpunct}{\mcitedefaultseppunct}\relax
\EndOfBibitem
\bibitem[Chau and Hardwick(1998)]{MP_1998_93_0511}
P.-L. Chau and A.~J. Hardwick, \emph{Molec. Phys.}, 1998, \textbf{93},
  511\relax
\mciteBstWouldAddEndPuncttrue
\mciteSetBstMidEndSepPunct{\mcitedefaultmidpunct}
{\mcitedefaultendpunct}{\mcitedefaultseppunct}\relax
\EndOfBibitem
\bibitem[Errington and Debenedetti(2001)]{N_2001_409_00318}
J.~R. Errington and P.~G. Debenedetti, \emph{Nature}, 2001, \textbf{409},
  318\relax
\mciteBstWouldAddEndPuncttrue
\mciteSetBstMidEndSepPunct{\mcitedefaultmidpunct}
{\mcitedefaultendpunct}{\mcitedefaultseppunct}\relax
\EndOfBibitem
\bibitem[Agarwal \emph{et~al.}(2011)Agarwal, Alam, and
  Chakravarty]{JPC_B_2011_115_06935}
M.~Agarwal, M.~P. Alam and C.~Chakravarty, \emph{J. Phys. Chem. B}, 2011,
  \textbf{115}, 6935\relax
\mciteBstWouldAddEndPuncttrue
\mciteSetBstMidEndSepPunct{\mcitedefaultmidpunct}
{\mcitedefaultendpunct}{\mcitedefaultseppunct}\relax
\EndOfBibitem
\bibitem[Santra \emph{et~al.}(2011)Santra, Klimes, Alfe, Tkatchenko, Slater,
  Michaelides, Car, and Scheffler]{PRL_2011_107_185701}
B.~Santra, J.~Klimes, D.~Alfe, A.~Tkatchenko, B.~Slater, A.~Michaelides, R.~Car
  and M.~Scheffler, \emph{Phys. Rev. Lett.}, 2011, \textbf{107}, 185701\relax
\mciteBstWouldAddEndPuncttrue
\mciteSetBstMidEndSepPunct{\mcitedefaultmidpunct}
{\mcitedefaultendpunct}{\mcitedefaultseppunct}\relax
\EndOfBibitem
\bibitem[Mishima and Endo(1980)]{JCP_1980_73_02454}
O.~Mishima and S.~Endo, \emph{J. Chem. Phys.}, 1980, \textbf{73}, 2454\relax
\mciteBstWouldAddEndPuncttrue
\mciteSetBstMidEndSepPunct{\mcitedefaultmidpunct}
{\mcitedefaultendpunct}{\mcitedefaultseppunct}\relax
\EndOfBibitem
\bibitem[Kamb and Prakash(1968)]{ACSB_1968_24_1317_nolotengo}
B.~Kamb and A.~Prakash, \emph{Acta Crystallographica Section B Structural
  Science}, 1968, \textbf{24}, 1317\relax
\mciteBstWouldAddEndPuncttrue
\mciteSetBstMidEndSepPunct{\mcitedefaultmidpunct}
{\mcitedefaultendpunct}{\mcitedefaultseppunct}\relax
\EndOfBibitem
\bibitem[Fortes \emph{et~al.}(2005)Fortes, Wood, Alfredsson, Vocadlo, and
  Knight]{JAC_2005_38_0612}
A.~D. Fortes, I.~G. Wood, M.~Alfredsson, L.~Vocadlo and K.~S. Knight, \emph{J.
  App. Crystallography}, 2005, \textbf{38}, 612\relax
\mciteBstWouldAddEndPuncttrue
\mciteSetBstMidEndSepPunct{\mcitedefaultmidpunct}
{\mcitedefaultendpunct}{\mcitedefaultseppunct}\relax
\EndOfBibitem
\bibitem[Saul and Wagner(1989)]{JPCRD_1989_18_1537}
A.~Saul and W.~Wagner, \emph{J. Phys. Chem. Ref. Data}, 1989, \textbf{18},
  1537\relax
\mciteBstWouldAddEndPuncttrue
\mciteSetBstMidEndSepPunct{\mcitedefaultmidpunct}
{\mcitedefaultendpunct}{\mcitedefaultseppunct}\relax
\EndOfBibitem
\bibitem[Fortes \emph{et~al.}(2010)Fortes, Wood, Norman, and
  Tucker]{ISIS_RB_1010211}
A.~D. Fortes, I.~G. Wood, L.~H. Norman and M.~G. Tucker, \emph{ISIS
  Experimental Report}, 2010, \textbf{1010211}, \relax
\mciteBstWouldAddEndPuncttrue
\mciteSetBstMidEndSepPunct{\mcitedefaultmidpunct}
{\mcitedefaultendpunct}{\mcitedefaultseppunct}\relax
\EndOfBibitem
\end{mcitethebibliography}
\bibliographystyle{rsc} 
}

\end{document}